\documentclass{article}

\usepackage[utf8]{inputenc}
\usepackage[T1]{fontenc}
\usepackage[english]{babel}
\usepackage{graphicx}
\usepackage[top=3cm, bottom=2cm, left=3cm, right=3cm]{geometry}
\usepackage[font=footnotesize]{caption}
\usepackage{amssymb}
\usepackage{amsmath}
\usepackage{titling}
\usepackage{setspace}
\usepackage{array}
\usepackage[usenames, dvipsnames]{color}
\usepackage{epstopdf}

\onehalfspace

\title{\bf{Percolation in suspensions of polydisperse hard rods: quasi universality and finite-size effects}}

\author
{Hugues Meyer$^{1}$, Paul van der Schoot $^{2,3}$, Tanja Schilling$^{1}$,\\
\\
\it
\normalsize{$^{1}$Research Unit for Physics and Materials Science, Universit\'{e} du Luxembourg,}\\
\it
\normalsize{L-1511 Luxembourg, Luxembourg}\\
\it
\normalsize{$^{2}$ Department of Applied Physics, Eindhoven University of Technology,} \\
\it
\normalsize{P.O. Box 513, 3500 MB Eindhoven, The Netherlands} \\
\it
\normalsize{$^{3}$ Institute for Theoretical Physics, Utrecht University,} \\
\it
\normalsize{Leuvenlaan 4, 3584 CE Utrecht, The Netherlands}
}

\date{}

\begin{document}

\maketitle

\begin{center}
  \large{\bf{Abstract}}\\
 \end{center}
We present a study of connectivity percolation in suspensions of hard spherocylinders by means of Monte Carlo simulation and connectedness percolation theory. We focus attention on polydispersity in the length, the diameter and the connectedness criterion, and invoke bimodal, Gaussian and Weibull distributions for these. The main finding from our simulations is that the percolation threshold shows quasi universal behaviour, i.e., to a good approximation it depends only on certain cumulants of the full size and connectivity distribution. Our connectedness percolation theory hinges on a Lee-Parsons type of closure recently put forward that improves upon the often-used second virial approximation \cite{schilling:2015}. The theory predicts exact universality. Theory and simulation agree quantitatively for aspect ratios in excess of 20, if we include the connectivity range in our definition of the aspect ratio of the particles. We further discuss the mechanism of cluster growth that, remarkably, differs between systems that are polydisperse in length and in width, and exhibits non-universal aspects.
\clearpage

\section{Introduction}
Composite nanomaterials have long attracted attention because of their potential application for instance in electronics, display technology and photovoltaics \cite{park:2013}.
Of particular interest in this context are their heat and charge transport properties \cite{li:2011}.
Adding  a sufficient amount of electrically conductive nanoparticles, such as carbon nanotubes or graphene, to an insulating polymer matrix produces a conductive composite the conductivity of which can be tuned by the choice of filler type, filler loading and processing \cite{ghislandi:2012}. For many technological applications, the minimum filler loading required to reach a conductive state, the so-called percolation threshold, is desired to be as low as possible \cite{torquato:2002}. Rod-like particles are particularly suitable for this kind of application since they present very low percolation thresholds \cite{koning:2012}. For the purpose of the rational design of such materials it is crucial to be able to describe and predict the percolation threshold of assemblies of filler particles, and understand the underlying mechanisms of the buildup of the system-spanning network required for effective conduction.

In experimental reality the properties of the filler nanoparticles are not always well controlled \cite{tkalya:2014}. Indeed, they are usually chemically and otherwise polydisperse, that is, consist of a mixture of particles of different dimensions and conductive properties. This complexity makes prediction of the percolation threshold and of the network structure very difficult, not least because of the huge parameter space. In this paper we present a simulation and theoretical study of percolation in dispersions of polydisperse nanorods, specifically allowing for hard core interactions and targeting aspect ratios that are of an intermediate range, i.e., not in the scaling limit \cite{otten:2009,otten:2011,nigro:2013}.

Even though we find qualitative agreement with work on polydisperse ideal (penetrable) rods \cite{nigro:2013} and very long hard rods \cite{otten:2009,otten:2011} (showing that the percolation threshold obeys laws that within a good approximation depend only on a few moments of the full distributions functions) quantitatively our results are very different. In fact, we find strong deviations in the dependence of the percolation threshold on the appropriate measures for the mean aspect ratio and connectivity of the particles. Finally, we find that the network connectivity properties are affected differently by variabilities in length, diameter and connectivity criterion.

It is important to point out that the model systems that have been studied in the literature so far usually capture only one or a few aspects relevant to experimental reality. A very large focus is on the particle shape, attractive interactions and aspect ratio \cite{schilling:2009}. While there is a huge body of literature dealing with monodisperse systems, relatively little attention has been paid to polydisperse systems \cite{kyrylyuk:2008, yi:2004, wang:2003, leung:1991, chatterjee:2000, munson-mcgee:1991, celzard:1996, pike:1974, neda:1999, foygel:2005, mutiso:2013, chatterjee:2012, berhan:2007, schilling:2007, schilling:2009, ambrosetti:2010, vigolo:2005}. Recently, Chatterjee \cite{chatterjee:2008} and Otten and Van der Schoot \cite{otten:2009,otten:2011} have developed theories of continuum percolation that take polydispersity into account and predict universal scaling laws for the percolation threshold. These predictions have only to a small extent been tested numerically.

In a recent simulation study, Nigro and co-workers confirm that for hard and penetrable rods that are polydisperse only in length the percolation threshold depends only weakly on the exact shape of the length distribution \cite{nigro:2013}. A similar finding was obtained by Mutiso and collaborators for mutually penetrable length and width polydisperse rods \cite{mutiso:2012}. They also find that finite-aspect-ratio corrections on the predictions of Otten and Van der Schoot are quite significant up to aspect ratios of about one hundred \cite{otten:2009,otten:2011}.

Here we go considerably beyond the scope of earlier work, and report on simulation results for three different types of polydispersity that we investigate separately. The coupling between different kinds of polydispersity, predicted to be relevant for many experimental systems \cite{otten:2011}, is postponed to future work. We show that the different kinds of polydispersity exhibit non-trivial universal behaviour. We invoke a treatment of connectedness percolation theory of hard rods recently put forward by us \cite{schilling:2015}, which is aimed at predicting finite-aspect-ratio corrections rather than obtaining them phenomenologically from simulations, as was done in \cite{mutiso:2012}.

The remainder of this paper is arranged as follows. We present in section 2 the methods implemented in the Monte-Carlo simulations. Section 3 deals with the derivation of our version of connectivity percolation theory for polydisperse spherocylinders, including the Lee-Parsons approximation. Finally, we focus in section 4 on both numerical and theoretical results, first about the percolation thresholds and then about the cluster mechanisms. We end the paper with conclusions and a summary of the main findings in section 5.

\section{Simulation Methods}

\begin{figure}
 \begin{center}
  \includegraphics[width=.45\linewidth]{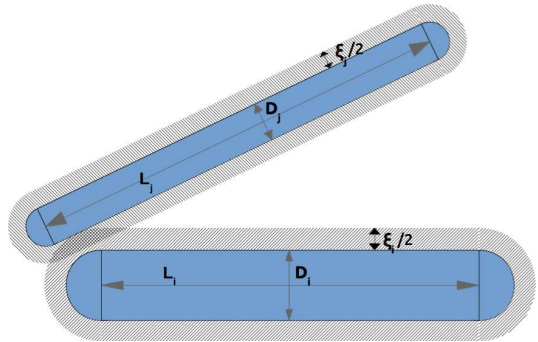}
 \end{center}
\caption{Definition of the particle dimensions and connectivity range.}
\label{sketch}
\end{figure}

We consider hard spherocylinders consisting of cylinders of length $L_{i}$ and diameter $D_{i}$, each capped by two hemispheres of the same diameter. See fig.~\ref{sketch}.
The particles are not allowed to overlap, but do not directly interact with each other when they are not in contact. The corresponding interaction potential is therefore either zero or infinite,
making their resulting equilibrium properties temperature independent.
We initialize a simulation box in which around $10,000$ spherocylinders are perfectly aligned and regularly placed on square lattices spaced from each other along the rod direction. 
At each simulation step, the particles are then randomly rotated and translated. Equilibration is monitored by computing the nematic order parameter, which is expected to reach a constant value at the equilibrium (0 in the isotropic phase).
Once the system is equilibrated, we generate ca.~5000 independent
configurations of the system
and average all quantities of interest over those configurations. In order to detect overlapping particles efficiently, the box is divided into a fine grid \cite{vink:2005}, where the unit cell length is chosen equal to the greatest rod diameter in the system, so that the computational cost increases linearly with the number of particles. This method is very fast but rather expensive in terms of memory.

We take polydispersity into account by assigning to each rod a length and a
diameter according to a
probability density function $\mathcal{P}_{L}$ and $\mathcal{P}_{D}$.
To generate a finite number of rods from a continuous distribution, we define
an interval $\Delta x$ where $x = L,D$ stands for length and width, and force the system to contain
$N\mathcal{P}_{x}(x_{0})\Delta x$
rods, whose dimension lies between $x_{0}$ and $x_{0}+\Delta x$.
This method turns out to give much more accurate results than simply drawing the rod dimensions directly from the distribution under study.
In this work, the mean aspect ratio lies around $L/D=15$
and the widest distributions we considered spread up to an aspect ratio of approximatively $L/D=80$ for the very longest rods. In order to clearly distinguish between the effects of length and diameter polydispersity we choose only uncorrelated distributions. As already advertised, this assumption does not necessarily apply to all experimental systems.

\begin{figure}
 \begin{center}
 \includegraphics[width=.45\linewidth]{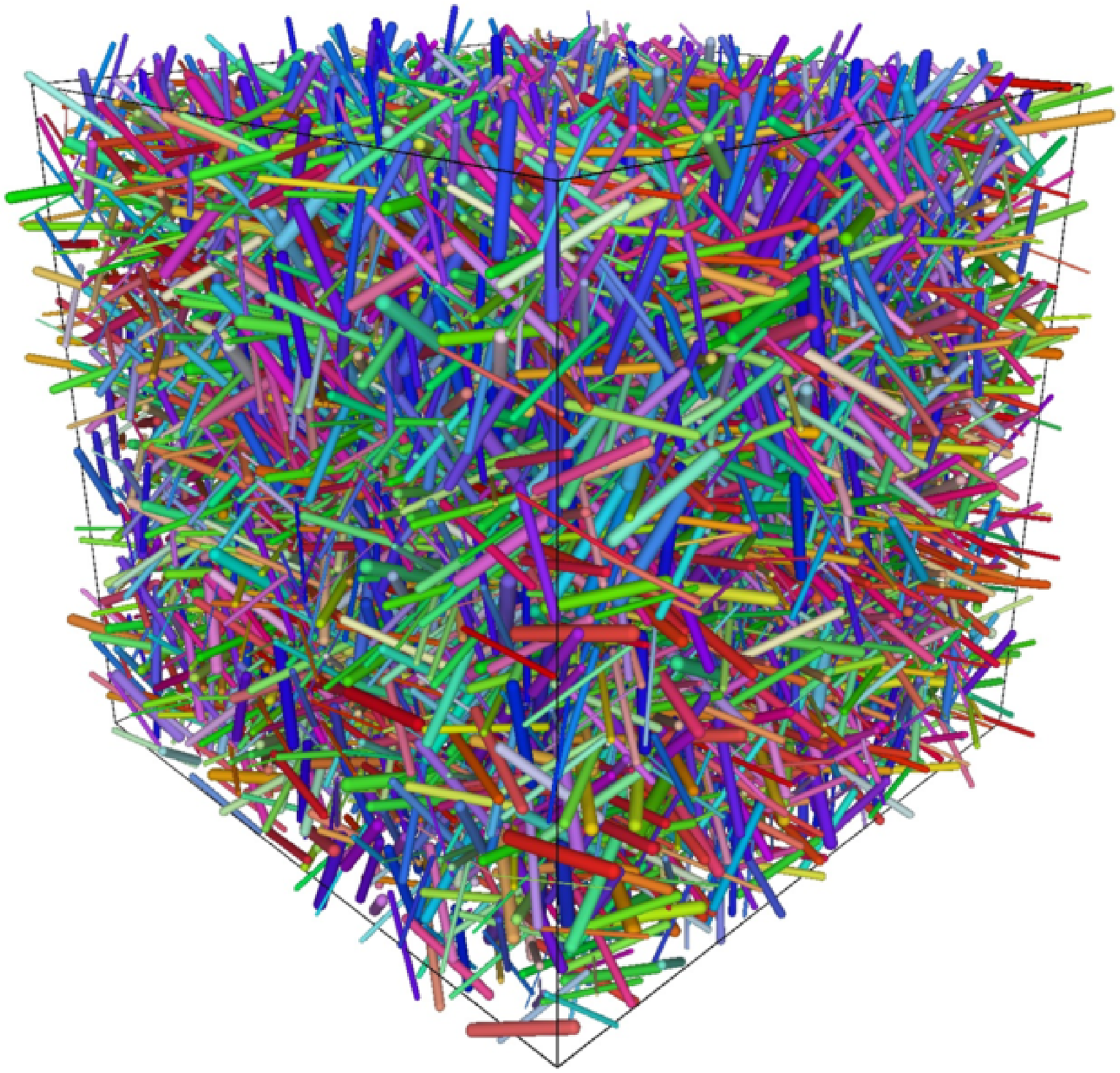}	
 	\hspace*{.04\linewidth}
  \includegraphics[width=.45\linewidth]{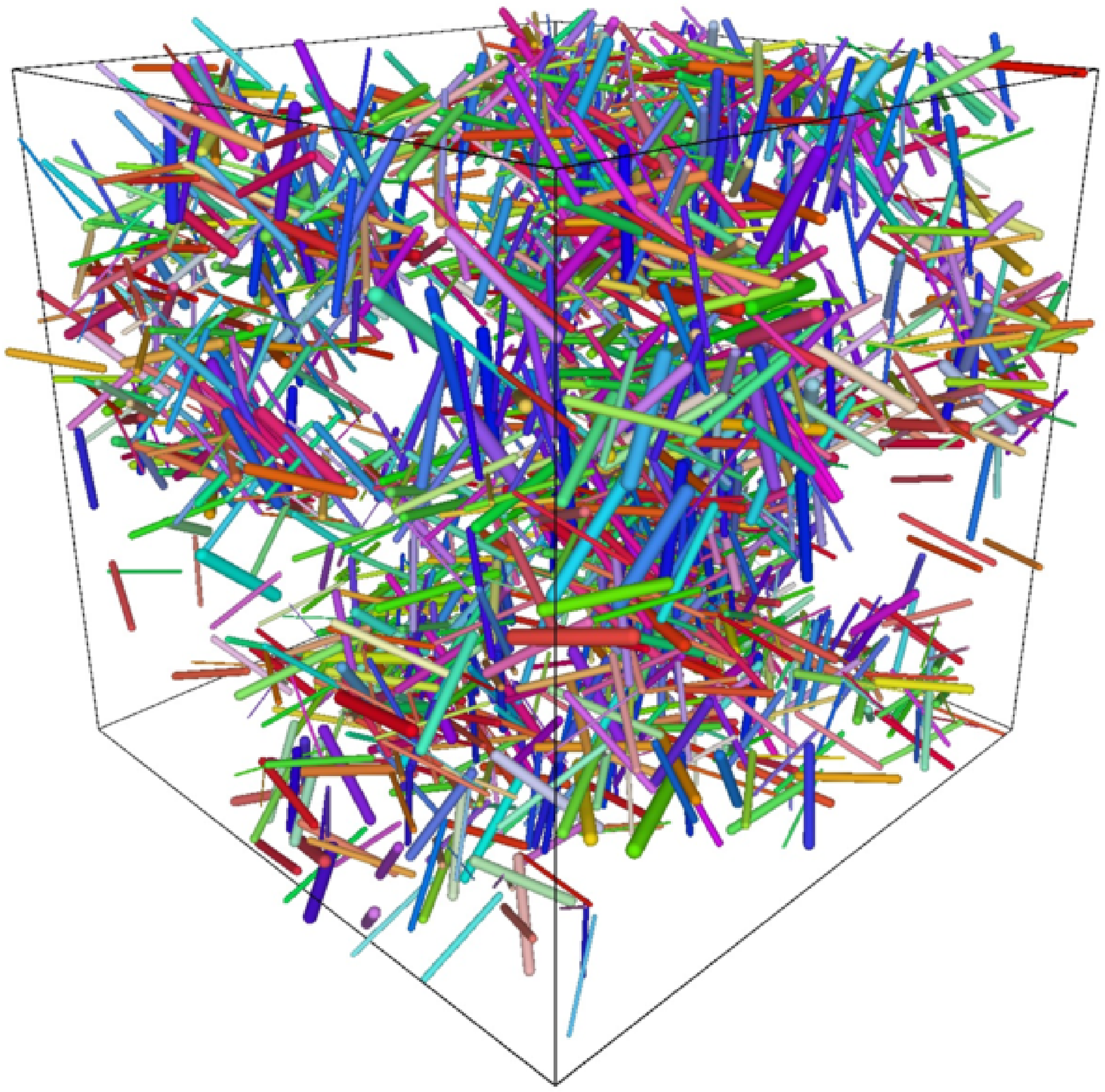}
\end{center}
  \caption{Snapshot of an equilibrated configuration of diameter polydisperse spherocylinders at the critical volume fraction (left) and of the largest cluster within this particular configuration (right). Distribution is of the Weibull form $P(D) \propto \left( \frac{D}{\beta} \right)^{\alpha-1} e^{-(D/\beta)^{\alpha} }$ with $\alpha=2.83$ and $\beta=1.16$. Lengths and connectendess distances are all fixed to $L=15$ and $\xi=0.2$.}
\label{snapshot}
\end{figure}

Connectedness percolation requires the definition of an inter-particle connectedness criterion. We define for each rod $i$ a spherocylindrical shell of length $L_{i}$ and diameter $D_{i}+\xi_{i}$ that contains the particle, where
the connectivity  parameter $\xi_{i}$ obeys some distribution function $\mathcal{P}(\xi)$.
Two particles are then connected if their surrounding shells overlap. Clusters are defined by contiguous pairwise connections.
We define a configuration percolating if one of its clusters is connected to its image under periodic boundary conditions.
To every configuration corresponds a percolation probability that is either 1 (it percolates) or 0 (it does not percolate),
and averaging over many configurations we compute a global continuous percolation probability for a particular system. 
A typical snapshot of such a sample and of its corresponding largest cluster is shown in fig. \ref{snapshot}.

In order to estimate the percolation threshold for a particular length, diameter or connectivity distribution, we perform simulations using this distribution for a range of rod volume fractions. The volume fraction $\phi$ is defined with respect to the hard core volume of the particles
and does not take into account the connectivity shell:
$\phi = \frac{1}{a^{3}}\sum\limits_{i=1}^{N}{v_{i}}$
where $v_{i}=\frac{\pi}{4}L_{i}D_{i}^{2}+\frac{\pi}{6}D_{i}^{3}$ is the volume of the particle $i$
and $a$ is the simulation box length.
The percolation probability in a finite system is a sigmoidal function of the volume fraction running from 0 to 1.
Its transition steepness increases with the box size
and reaches a Heaviside step function in the limit of infinite box volume.
The curves that correspond to different box volumes cross each other slightly below the concentration at which the probability reaches the value of $0.5$. As we are interested in the scaling behaviour of the percolation threshold with the aspect ratio and cumulants of the size distribution of the particles, we do not need very accurate estimates. Hence,
we ignore finite size effects and assume that the percolation threshold is the volume fraction corresponding to a percolation probability of $0.5$. 
We verify that our box is sufficiently large to ensure that the percolation probability goes from 0.2 to 0.8 within a maximal volume fraction range $\Delta\phi_{max} = 0.005$. We assume this criterion to be restrictive enough for our results to achieve a satisfactory accurracy.

\section{Theory}
Percolation of clusters of nanoparticles in a fluid background medium can be investigated theoretically invoking what in essence is liquid state integral equation theory \cite{torquato:2002}. The theoretical framework is called connectedness percolation theory and it has been applied to hard and soft rod-like particles \cite{otten:2011,otten:2009,kyrylyuk:2008}.
Here, we follow the same recipe, except that we will not rely on the second virial approximation that becomes exact in the limit of infinite aspect ratio. Instead we opt for a closure that is was recently shown to provide an accurate description of percolation of monodisperse, hard rods with an aspect ratio larger than roughly ten \cite{schilling:2015}.

Within the framework of connectedness percolation theory, the cluster size $S$ can be expressed in terms of a function $T$, averaged over all of the attributes of orientation vector $\mathbf{u}$ and dimensions ${\mathbf x} \equiv (L,D,\xi)$ of the particles \cite{otten:2011},
\begin{equation}
 S = \left< T({\bf x},{\bf u}) \right> _{{\bf x},{\bf u}} \quad .
\end{equation}
The function $T$ itself is a pair connectedness function averaged over its attributes, and the solution of a generalised connectedness Ornstein-Zernike equation
\begin{equation}
    T({\bf x},{\bf u}) - \rho \left< \hat{C}^{+}(0,{\bf x}, {\bf x}',{\bf u'}) T({\bf x}', {\bf u'}) \right>_{{\bf x}', {\bf u'}} = 1 \quad .
\label{main_eq}
\end{equation}
Here, $\rho$ is the number density of particles and $\hat{C}^{+}=\hat{C}^{+}({\mathbf q}, {\mathbf u}, {\mathbf u}')$ the spatial Fourier transform of the \textit{connectedness} direct correlation function, that is,
the direct correlation function for particles that are part of
the same cluster, and ${\mathbf q}$ the wave vector. To average over the entire volume of the system, we have to take the zero wave vector limit, ${\mathbf q}\rightarrow 0$.

In the isotropic phase the rods are randomly oriented, implying that $T({\mathbf x},{\mathbf u})=T({\mathbf x})$, which in turn allows us to redefine $\hat{C}^{+}$ as its average over the possible orientations, producing the simplified connectedness Ornstein-Zernike equation
\begin{equation}
 T({\bf x}) - \rho \left< \hat{C}^{+}(0,{\bf x}, {\bf x}') T({\bf x}') \right>_{{\bf x}'} = 1
 \label{main_isotropic}
\end{equation}

This equation needs to be closed and we follow Schilling et al. \cite{schilling:2015} by invoking the following \textit{Ansatz}:  $\hat{C}^{+}(0,{\bf x},{\bf x}') = \Gamma(\phi)\hat{f}^{+}(0,{\bf x},{\bf x}')$ \cite{schilling:2015}.
Here, $f^{+} = e^{-\beta u^{+}}$ is the connectedness Mayer function and $\hat{f^{+}}$ its spatial Fourier tranform  \cite{otten:2011}, with $\beta$ the reciprocal thermal energy and $u^{+}$ the so-called connectedness potential, and $\Gamma(\phi)$ a coefficient that depends on the volume fraction $\phi$ of the particles. The functional form of $\Gamma$ is obtained from the Lee-Parsons expression for the excess free energy, interpolating between the Percus-Yevick equation of state for hard spheres and the Onsager equation of state for hard rods \cite{parsons:1979,lee:1987, percus:1958, thiele:1963,cinacchi:2002}.
Within this \textit{Ansatz}, we have
\begin{equation}
 \Gamma(\phi) = \frac{1-\frac{3}{4}\phi}{(1-\phi)^{2}}.
\end{equation}
Notice that in the limit $\phi\rightarrow 0$, $\Gamma\rightarrow 1$ and we obtain the second virial theory that is valid in the Onsager limit of very slender rods. As we demonstrated recently \cite{schilling:2015}, corrections to the Onsager limit are significant for aspect ratios below a few hundred.

For the case of hard rods, the connectedness Mayer function is 1 if the distance between two rods is between $D$ and $\Delta\equiv D+\xi$
and 0 otherwise. The zero-wave vector Fourier transform of the connectedness Mayer function $\hat{f}^{+}(0,{\bf x},{\bf x}')$ can be separated into contributions from interactions between the different portions of the spherocylindrical particles. We use Onsager's expression for the excluded volume of two hard spherocylinders to obtain \cite{onsager:1949}
\begin{equation}
\begin{split}
 \hat{f}^{+}(0,{\bf x},{\bf x}') = LL'\frac{\Delta+\Delta'}{2}f_{11} + (L+L')\left( \frac{\Delta+\Delta'}{2} \right)^{2}f_{10} + \left( \frac{\Delta+\Delta'}{2} \right)^{3}f_{00}\\
 - LL'\frac{D+D'}{2}f_{11} - (L+L')\left( \frac{D+D'}{2} \right)^{2}f_{10} - \left( \frac{D+D'}{2} \right)^{3}f_{00}
\end{split}
 \label{maier}
\end{equation}
where $\Delta = D+\xi$ and the coefficients $f_{00}=4\pi/3$, $f_{10}=\pi$ and $f_{11}=\pi/2$ respectively
indicate the cylinder-cylinder, cylinder-hemisphere and hemisphere-hemisphere contributions. Note that because $u^{+}$ is infinite for particles with overlapping hard cores and for those that are not connected, and zero for connected ones, $\hat{f}^{+}$ is essentially the difference between the excluded volume of two particles with hard-core radius $\Delta$ minus that of particles with hard-core radius $D$.

Let us first focus on length polydispersity alone, and set $D=D'$ and $\Delta = \Delta'$. In this particular case eq.~\ref{maier} becomes
\begin{equation}
 \hat{f}^{+}(L,L') = LL'\omega_{1}f_{11} + (L+L')\omega_{2}f_{10} + \omega_{3}f_{00},
\end{equation}
where $\omega_{n}=\Delta^{n}-D^{n}$ are differences between powers of the interaction ranges $\Delta$ and $D$. Inserting this into equation \ref{main_isotropic}, we find
\begin{equation}
 T(L) - \rho \Gamma(\phi) (L\omega_{1}f_{11} + \omega_{2}f_{10}) \left< LT(L) \right>_{L} -
 \rho \Gamma(\phi) (L\omega_{2}f_{10} + \omega_{3}f_{00}) \left< T(L) \right>_{L} = 1.
 \label{main_dev}
\end{equation}
The two unknown coupled quantities are $\left< T(L) \right>_{L}$ and $\left< LT(L) \right>_{L}$,
therefore we need two independent equations relating them.
The first one we obtain by averaging equation \ref{main_dev},
whereas the second we derive by first multiplying equation \ref{main_dev} with $L$ and averaging the resulting equation.
As these relations are linear, we can summarize them by
defining two vectors $\underline{X}$ and $\underline{Y}$ and a matrix $\underline{\underline{M}}$:
\begin{equation}
   \underline{X}=\left[
  \begin{tabular}{c}
    $\left< T(L) \right>$\\
    $\left< LT(L) \right>$
   \end{tabular}
  \right] \quad , \quad
  \underline{Y} = \left[
  \begin{tabular}{c}
    $1$\\
    $\left< L \right>$
   \end{tabular}
  \right] \quad , \quad
  \underline{\underline{M}} = \left[
  \begin{tabular}{c c}
    $1-k\alpha_{1}$ & $-k\beta_{1}$ \\
    $-k\alpha_{2}$ & $1-k\beta_{2}$
   \end{tabular}
  \right] \quad , \quad
   \underline{\underline{M}} \underline{X}=\underline{Y}
\end{equation}
where we use the notation $k \equiv \rho\Gamma(\phi)$,
$\alpha_{1} \equiv \left< L \right> \omega_{2}f_{10} + \omega_{3}f_{00}$,
$\alpha_{2} \equiv \left< L^{2} \right> \omega_{2} f_{10} + \left< L \right> \omega_{3}f_{00}$,
$\beta_{1} \equiv \left< L \right> \omega_{1}f_{11} + \omega_{2}f_{10}$, and
$\beta_{2} \equiv \left< L^{2} \right> \omega_{1}f_{11} + \left< L \right> \omega_{2}f_{10}$. The cluster size $S=\left< T(L) \right>$ is the first element of $\underline{X}=\underline{\underline{M}}^{-1}\underline{Y}$.
Each element of the matrix $\underline{\underline{M}}^{-1}$ is a fraction
whose denominator is the determinant of $\underline{\underline{M}}$.

Therefore, $S$ can be written as
\begin{equation}
 S=\frac{\sum\limits_{n,p,q}{A_{npq} \left< L^{n} \right> D^{p} \xi^{q}}}{\det{M}}
\end{equation}
where $A_{npq}$ are coefficients which involve the quantities $k$ and $f_{ij}$. However, their exact expressions are not important for the remaining calculations, so we do not reproduce them here. The percolation threshold is the volume fraction $\phi_{c}$
for which $S$ diverges, or, equivalently, that makes
$\det{\underline{\underline{M}}}=1-k(\alpha_{1}+\beta_{2})+k^{2}(\alpha_{1}\beta_{2}-\alpha_{2}\beta_{1})$ vanish.
Therefore, we determine $\phi_{c}$ by solving this simple second-order polynomial equation for $k$
and by then relating $k$ and $\phi$, $ \phi\Gamma(\phi) = k \left< v \right>
$, where $\left< v \right> = \frac{\pi}{4} \left( \left< L \right> D^{2} + \frac{2}{3} D^{3} \right)$
is the average volume of a particle.
The polynomial equation yields two solutions $k_{\pm}$. We define then $\gamma_{\pm}=k_{\pm}\left< v \right>$.

The final equation that we have to solve is therefore
\begin{equation}
 \left( \gamma_{\pm} + \frac{3}{4} \right)\phi_{c}^{2} - (1+2\gamma_{\pm})\phi_{c} + \gamma_{\pm} = 0,
\end{equation}
which has two solutions for $\phi_{c}$,
\begin{equation}
 \phi_{c} = \frac{1+2\gamma_{\pm}\pm\sqrt{1+\gamma_{\pm}}}{2\gamma_{\pm}+\frac{3}{2}}.
\end{equation}
We have four solutions for $\phi_{c}$ but we only keep the positive solution below unity for obvious physical reasons.

The analysis of the cases of diameter and connectedness polydispersity are completely analogous to that of length polydispersity, that is,
the same method is applied starting from eq.~\ref{maier} setting $L=L'$ and $\xi=\xi'$, and $L=L'$ and $D=D'$, respectively. $\underline{\underline{M}}$ then turns into a $3\times3$ matrix for diameter polydispersity
and a $4\times4$ one for connectedness polydispersity. This leads to third and fourth order polynomial equations
that need to be solved. It is important to note that the percolation threshold will then explicitly depend on the higher order moments $\left< D^{n} \right>$
with $n\leq4$ and $\left< \xi^{m} \right>$ with $m \leq 6$, respectively.

\section{Results}

\subsection{Percolation thresholds}

If we invoke the second-virial approximation and neglect the end-cylinder and end-end interactions, the percolation threshold, $\phi_c$, becomes proportional to the reciprocal weight average length
$\left< L \right>_{w}^{-1} \equiv \left< L \right> / \left< L^{2} \right>$, the mean square width $\left< D^{2} \right>$ and a measure for the mean reciprocal connectivity length
$\left[\left< \xi \right> + \sqrt{\left< \xi^{2} \right>}\right]^{-1}$, depending on the type of polydispersity \cite{otten:2011,otten:2009,nigro:2013}.
Although this approximation produces results that are not very accurate
for aspect ratios that are not huge, it is useful to take it as a reference because it shows
what cumulants of the full distribution are expected to govern the percolation threshold.
We note in this context that the nature of length and diameter polydispersity is fundamentally different from that of connectedness distance polydispersity. The first two relate to polydispersity in the particle dimensions and hence in the interactions between particles,
whilst the third one is a polydispersity in the electrical connectivity length scale only, which does \textit{not} affect the structure of the liquid but only the resulting cluster size distribution.

To separate these various effects, we define two new dimensionless quantities, being
$\chi=\left< L \right>_{w}/\sqrt{\left< D^{2} \right>}$ and
$\lambda=\left[\left< \xi \right> + \sqrt{\left< \xi^{2} \right>}\right]/2\sqrt{\left< D^{2} \right>}$. The former,
$\chi$, becomes equal to either $\left< L \right>_{w}/D$ or $L/\sqrt{\left< D^{2} \right>}$ depending on the type of polydispersity
and is analogous to the aspect ratio, as we can not define a unique aspect ratio in polydisperse systems. The latter,
$\lambda$, becomes $\left[\left< \xi \right> + \sqrt{\left< \xi^{2} \right>}\right]/2D$ for rods with a monodisperse diameter,
and represents a characteristic connectedness shell thickness compared to the particle diameter.

We compare the percolation threshold obtained from our simulations, prediction based on the theory presented in the previous section
and those from the second-virial  theory for which $\Gamma=1$, as a function of $\chi^{-1}$ for length and diameter polydispersity
and as a function of $\lambda^{-1}$ for connectedness polydispersity.

For all three kinds of polydispersity, we tested bidisperse, Gaussian and Weibull distributions.
The first describes binary mixtures,
the second seems relevant as Gaussian distributions are common
in many fields of physics,
and the third has been experimentally observed in polymer-fiber composites that are polydisperse in length \cite{wang:2006,hine:2002}.

\begin{figure}
 \begin{center}
  \includegraphics[width=.45\linewidth]{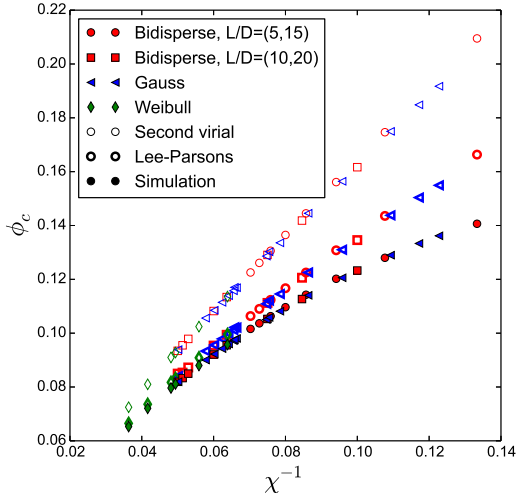}
 \end{center}
\caption{Percolation threshold $\phi_c$ for length polydispersity as a function of the reciprocal aspect ratio $\chi^{-1}=D/\left< L \right>_{w}$. Results are indicated from simulations (full dots), Lee-Parsons theory as well as the second-virial approximation (empty dots) for $\xi/D=0.2$ and various distributions $\mathcal{P}(L)$. A remarkable universal scaling with $\chi^{-1}$ is observed in the three cases. See also the main text.}
\label{threshold:length}
\end{figure}

\begin{figure}
 \begin{center}
  \includegraphics[width=.45\linewidth]{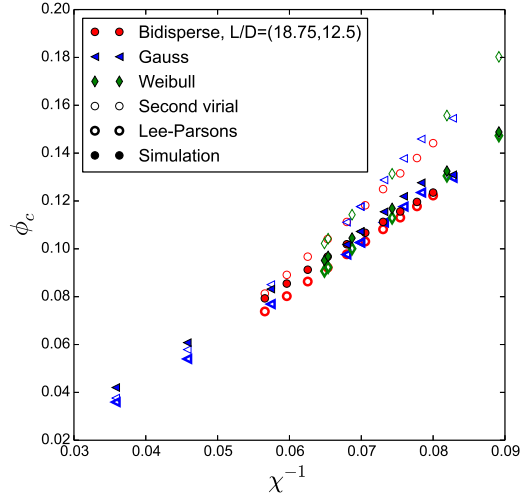}
 \end{center}
\caption{Percolation threshold $\phi_c$ for diameter polydispersity as a function of $\chi^{-1}=\sqrt{\left< D^{2} \right>}/L$. Results are indicated from our simulations (full dots), Lee-Parsons theory as well as the second-virial approximation (empty dots) for $L/\xi=75$ and various distributions $\mathcal{P}(D)$.
  The scaling with $\chi^{-1}$ is not anymore universal, higher order cumulants matter.  See also the main text.}
\label{threshold:diameter}
\end{figure}

\begin{figure}
 \begin{center}
  \includegraphics[width=.45\linewidth]{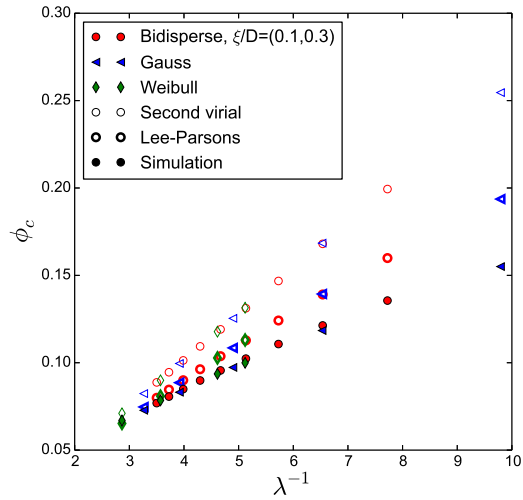}
 \end{center}
\caption{Percolation threshold $\phi_c$ for polydispersity in the connectedness distance $\xi$, as a function of $\lambda^{-1}=2D/\left[\left< \xi \right> + \sqrt{\left< \xi^{2} \right>}\right]$. Results from simulations (full dots), Lee-Parsons theory as well as second-virial approximation (empty dots) for $L/D=15$ and various distributions $\mathcal{P}(\xi)$. A universal scaling with $\lambda^{-1}$ is also obtained.
   See also the main text.}
\label{threshold:chemical}
\end{figure}

Fig.~\ref{threshold:length} shows the percolation threshold for the case of length polydispersity as a function of the inverse aspect ratio $\chi^{-1}$. Both theory and simulations display a remarkable
universal behaviour of the percolation threshold as a function of
$\left< L \right>_{w}$.
The three different distributions that we tested are very different in shape, but we find that the percolation threshold is not sensitive to this if expressed in terms of the weight average length $\left< L \right>_{w}$.
This was also shown by Nigro et al. \cite{nigro:2013} for penetrable particles and in a more limited fashion for hard particles.
We find that the Lee-Parsons theory and our simulation results quantitatively converge in the range $\chi > 20$ and are in qualitative agreement below that,
whereas the theoretical prediction derived from the second virial approximation deviates notably even for relatively large aspect ratios.
For monodisperse rods, the Lee-Parsons approach produces quantitative
results already for $\chi > 10$ \cite{schilling:2015}.
In polydisperse systems, the discrepancies between theory and simulations
extend to larger average aspect ratios because of the shorter rods that
are also present in the system. Note that the simulation results are always below the theoretical prediction. Hence, composite materials that contain fibres of short aspect ratio do not need as high filler loadings as theoretically expected in order to become conductive.

Also for diameter polydispersity the
second order cumulant is expected to be the most relevant one
according to the second-virial theory, at least if we neglect
end effects \cite{otten:2011}.
This is confirmed in fig.~\ref{threshold:diameter} by both our simulation results and the more accurate predictions of Lee-Parsons theory
albeit that we do not observe strictly universal behaviour. This suggests that higher order cumulants must be important too, at least for aspect ratios below $25$.
Remarkably, even the second-virial prediction yields results that are in almost quantitative agreement with the simulations for $\chi>15$.
The Lee-Parsons theory is in very good agreement with the simulation results even for relatively short particles.
We note that this theory underestimates the percolation threshold for
diameter polydisperse spherocylinders
while it overestimates it for length polydisperse ones.

Finally, we focus on connectedness distance polydispersity, illustrated in fig.~\ref{threshold:chemical} for the case of an aspect ratio of $15$.
As explained above, this does not lead to variability in the
interactions between the particles, only in the definition of which particles are part of the same cluster.
Nevertheless, it can be treated theoretically in the same way as the length and the diameter polydispersity due to the definition of the connectedness potetial $u^{+}$.
This explains why we find similar behaviour as for length and diameter
polydispersity. Indeed, there is quasi-universal scaling with respect to
$\lambda$, even though the aspect ratio is not all that large and one would expect higher order moments in the distribution of connectedness ranges to show up.
Indeed, our calculations show contributions up to the sixth moment, $\left< \xi^{6} \right>$, for the cluster size.
Apparently terms involving the first and second moments, $\left< \xi \right>$ and $\left< \xi^{2} \right>$, predominate the percolation threshold.

Again, prediction from the second-virial approximation disagrees significantly with the simulations. The Lee-Parsons correction improves upon the quality of the prediction, but still overestimates the threshold.
Both theories improve for large values of $\lambda$, i.e., for globally thick connectedness shells for which the percolation threshold occurs at low volume fractions of particles.
Our \textit{Ansatz} for the connectedness direct correlation function has by construction the spatial structure of a second virial theory, even though that our Lee-Parsons extension does have the thermodynamics that goes beyond it. Arguably, at higher densities the actual structure of the direct correlation function starts to deviate from this. Note also that the Percus-Yevick prediction for the percolation threshold of monodisperse hard spheres is anyway in only qualitative agreement with simulations \cite{desimone:1986}.

\subsection{Cluster formation mechanisms}
\begin{figure}
 \begin{center}
  \includegraphics[width=.45\linewidth]{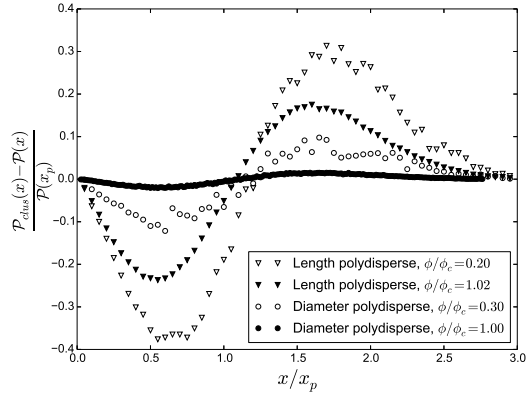}
 \end{center}
\caption{Difference between the length and diameter distribution $P_{cluster}(x)$ of the largest cluster and the distribution $P(x)$ of the whole system, normalized by the value of the distribution at its peak, $P(x_p)$, as a function of the length $x=L$ or width $x=D$ scaled to the peak value, for two volume fractions in both cases. The length (triangles) and diameter (circles) distributions of the entire system are of the Weibull form $P(x) \propto \left( \frac{x}{\beta} \right)^{\alpha-1} e^{-(x/\beta)^{\alpha} }$ with $\alpha=2.37$ and $\beta$ being such that the distribution peaks lie at $L=15$ and $D=1$. Larger and thicker particles cluster more easily. The difference between the distributions within clusters and the global ones become smaller with increasing volume fraction.
Notice that length polydispersity has a much stronger fractionation effect than diameter polydispersity.}
\label{mechanism:1}
\end{figure}

\begin{figure}
 \begin{center}
  \includegraphics[width=.45\linewidth]{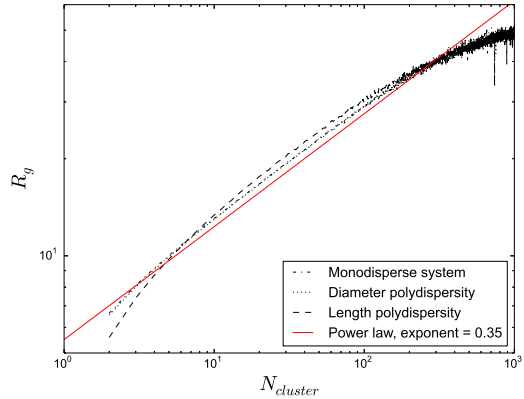}
 \end{center}
  \caption{Radius of gyration of clusters as a function of their size
  on a logarithmic scale for hard rods. Compared are results for monodisperse and for length and width polydisperse rods (see legend key). We set $D=1$ and $L=15$ in the monodisperse system. Length and width distributions are of the Weibull form  $P(x) \propto \left( \frac{x}{\beta} \right)^{\alpha-1} e^{-(x/\beta)^{\alpha} }$ with $\alpha=2$ and $\beta$ chosen such that $\left< L \right>_{w}=15$ and $\left< D^{2} \right>=1$ in both other systems. $\phi \simeq \phi_{c}$ and $\xi=0.2$ in all three cases. Monodisperse systems as well as systems polydisperse in diameter superimpose and can be fitted by a power law exhibiting an exponent around 0.35 (deviations from this scaling for very large clusters are finite-size effects). This is not true for length polydispersity.}
\label{mechanism:2}
\end{figure}

As we have seen, the sensitivity of the percolation threshold to
the higher order moments of the distribution function depends on
the type of polydispersity: length, diameter or connectedness range.
It suggests that the mechanism by which the particles cluster differs between the different types of polydispersity. Indeed, hard particles of different size and/or shape have for entropic reasons a tendency to phase separate, and even if they do not actually phase separate this might give rise to fractionation of particles in the transient clusters that form in the mixtures \cite{vanderschoot:1992, vanroij:1996}.

To investigate this, we compare in fig.~\ref{mechanism:1} the
distribution of lengths and diameters within the largest cluster
$\mathcal{P}_{\rm clus}(x)$ with that in the entire system
$\mathcal{P}(x)$ with $x=(L,D)$, for length and width polydisperse rods respectively. In both cases the larger particles are more abundant in the
largest cluster than in the whole system, explaining why relatively small amounts of large particles have a large effect on the percolation threshold \cite{otten:2009}. On the other hand, the effect weakens with increasing volume fraction of particles, at least for the length polydisperse ones. The proportion of short particles within large clusters is more and more important as packing fraction increases, making the gap between $\mathcal{P}_{\rm clus}(x)$ and $\mathcal{P}(x)$ smaller.

Another measure for the cluster structure is the fractal dimension $d_f$ of the critical cluster. We obtain this by measuring the radius of gyration $R_f \sim N_{\rm cluster}^{1/d_f}$ of the
clusters as a function of the number of particles in it, $N_{\rm cluster}$, see fig.~\ref{mechanism:2}. This quantity is sensitive to length polydispersity and less to diameter polydispersity. The latter and that for monodisperse rods collapse exactly on the same curve that seems to exhibit a power law scaling and a fractal dimension of $d_f=2.8$. Note that finite box-size effects cause deviations from pure power-law behaviour. For length polydisperse rods we also do not find power law scaling but the trend seems to conform to the same fractal dimension but with a larger prefactor. The fractal dimension of $2.8$ is larger than the mean-field value of $2$ we expect to hold for very long rods \cite{otten:2012} and that we obtain from the second virial approximation, but close to the accepted value of $2.5$ in three dimensions for standard percolation \cite{stauffer:1994}. We expect that because in our simulations the aspect ratio of the particles is not very large that we find a deviation from the mean-field exponent.

We conclude that having varying lengths or diameters within a collection of hard rods does not fundamentally change the way percolation is reached. Still, the volume fraction at the percolation threshold is different from
that for monodisperse rods, even if the average diameters (or lengths) are equal.

\section{Conclusions}
In summary, we have presented a theoretical and computer simulation
study on the effects of polydispersity on the geometrical percolation in suspensions of hard
spherocylinders. We compare results for bidisperse, Gaussian
and Weilbull distributions and show that the percolation threshold is quite
insensitive to the precise distribution. In the case of length and connectedness polydispersity the thresholds superpose within numerical error when scaled with the appropriate second order cumulant of the size distribution. For diameter polydispersity, however, higher order moments seem to matter, as the superposition of the different distributions is not quite perfect. To analyse the simulation results, we also present a theoretical treatment of the problem within connectedness percolation theory that we find to quantitatively predicts the percolation threshold for hard rods of aspect ratios above $20$.

\section*{Acknowledgements}
This project were completed within the framework of the ARPE program of the \'{E}cole Normale Sup\'{e}rieure de Cachan, France.
Data from computer simulations presented in this paper were carried out using the HPC facilities of University of Luxembourg.


\end{document}